\documentclass{iau}
\usepackage{graphicx}

\begin{document}

\lefttitle{M. Alizadeh et al.}
\righttitle{UKIRT M33 Variables, SFH, and Dust}

\jnlPage{1}{2}
\jnlDoiYr{2026}
\doival{10.1017/xxxxx}

\aopheadtitle{Proceedings IAU Symposium}
\editors{M. Zaja\v{c}ek, T. Je\v{r}\'{a}bkov\'{a}, V. Karas, R. Sch\"odel \& P. Sukov\'{a}, eds.}

\title{The UKIRT M33 Monitoring Project: A 15-Year Near-Infrared Variable
Star Catalogue for the Central Kiloparsec, and Prospects for Star
Formation History and Dust Return}

\author{Mina Alizadeh$^{1,2}$, Fatemeh Nikzat$^2$, Atefeh Javadi$^2$,
Yousefali Abedini$^{1,5,6}$, Jacco Th. van Loon$^3$ and Hedieh
Abdollahi$^{2,4}$}

\affiliation{$^1$Department of Physics, Faculty of Science,
University of Zanjan, 38791-45371, Zanjan, Iran
\email{minaalizadah@gmail.com}}
\affiliation{$^2$School of Astronomy, Institute for Research in
Fundamental Sciences (IPM), Tehran, 19568-36613, Iran
\email{atefeh@ipm.ir}}
\affiliation{$^3$Lennard-Jones Laboratories, Keele University,
Keele, Staffordshire ST5 5BG, UK}
\affiliation{$^4$Konkoly Observatory, HUN-REN Research Centre for
Astronomy and Earth Sciences, Budapest, Hungary}
\affiliation{$^5$Center for Research in Climate Change and Global Warming (CRCC), IASBS, Zanjan, Iran}
\affiliation{$^6$Research Center in Cancer Prevention Sciences and Technology, University of Zanjan, Zanjan, Iran}

\begin{abstract}
We present results from the UKIRT M33 Monitoring Project, a
near-infrared survey of variable red giants in the Local Group
spiral galaxy M33. Combining four independent photometric datasets
spanning UIST (2003), UFTI (2005), WFCAM (2005--2007), and a new
UKIRT Hemisphere Survey epoch (2018), we construct a homogenised
15.11-year $K$-band light-curve catalogue for 847 stars in the
central kiloparsec, of which 771 (91 per cent) are variable.
Cross-matching with archival Spitzer photometry identifies 120
dust-enshrouded AGB candidates. We compare the spatial coverage of
this central catalogue with the earlier UIST-only (Paper~I) and
disc-wide WFCAM (Paper~IV) variable-star surveys. Building on this
catalogue, two forthcoming papers will (i) measure individual
pulsation periods to reconstruct the star formation history of the
central kiloparsec, and (ii) refine dust and gas mass-loss rates
across the disc using the extended time baseline.
\end{abstract}


\maketitle

\section{The three surveys}

The UKIRT M33 monitoring project has grown through three complementary
surveys of increasing spatial reach. The programme began with UIST
near-infrared imaging of the central square kiloparsec of M33 between
2003 and 2007, from which \cite{Javadi2011a}
identified 812 variable stars, predominantly asymptotic giant branch
(AGB) stars, among 18\,398 monitored sources. The survey was later
extended by \cite{Javadi2015} to cover almost
the full extent of the galactic disc ($\sim$1~deg$^2$) using WFCAM,
yielding 4\,643 variable red giants from 403\,734 photometered stars.

In this work we return to the central field with a substantially
extended time baseline. We homogenise four independent near-infrared
photometric datasets onto a common photometric system: the original
UIST monitoring, UFTI and WFCAM-PSF imaging, and a single modern
epoch from the UKIRT Hemisphere Survey (UHS, 2018). This yields a
15.11-year $K$-band light-curve catalogue for 847 stars in the central
3-arcmin field, comprising 21\,085 individual measurements, of which
771 (91 per cent) are variable. The UHS epoch extends the UIST-era
baseline by over a decade, enabling long-period variability studies
not possible with UIST data alone.

\begin{figure}
\begin{center}
\includegraphics[width=2.75in]{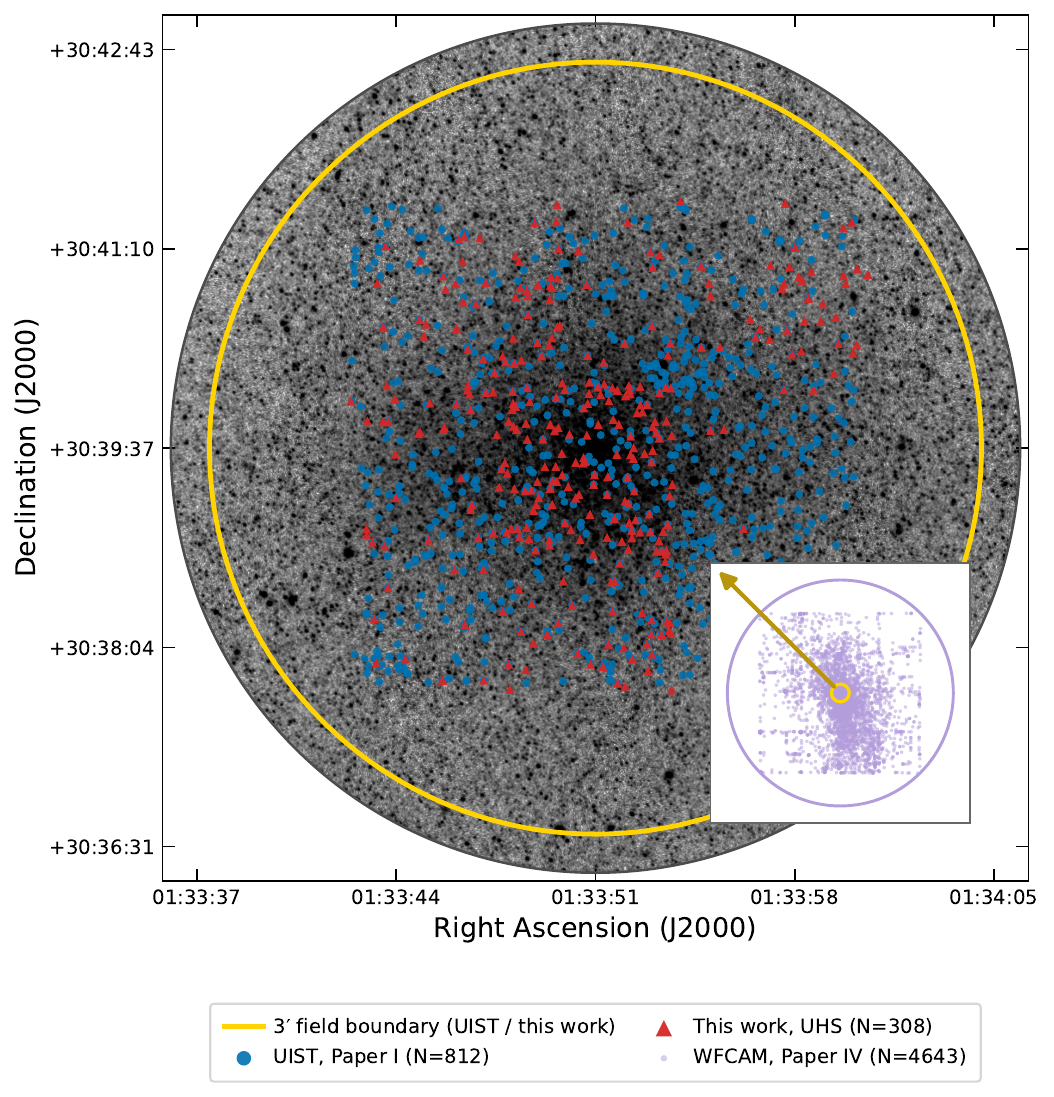}
\caption{Spatial footprint of the three near-infrared variable-star
surveys of M\,33. Main panel: UKIRT WFCAM $K$-band image of the
central field (single detector, this work), with axes taken directly
from the image World Coordinate System. The yellow circle marks the
3-arcmin-radius field common to the UIST (\cite{Javadi2011a}; 812
variables, circles) and UHS (this work; 308
variables, triangles) surveys. Inset: the disc-wide WFCAM survey
(\cite{Javadi2015}; 4643 variables, points)
extends out to $\sim$38 arcmin from the nucleus; the small yellow
circle shows the location and scale of the main panel within this
wider field.}
\label{fig:fields_iau}
\end{center}
\end{figure}

\begin{figure}
\begin{center}
\includegraphics[width=2.75in]{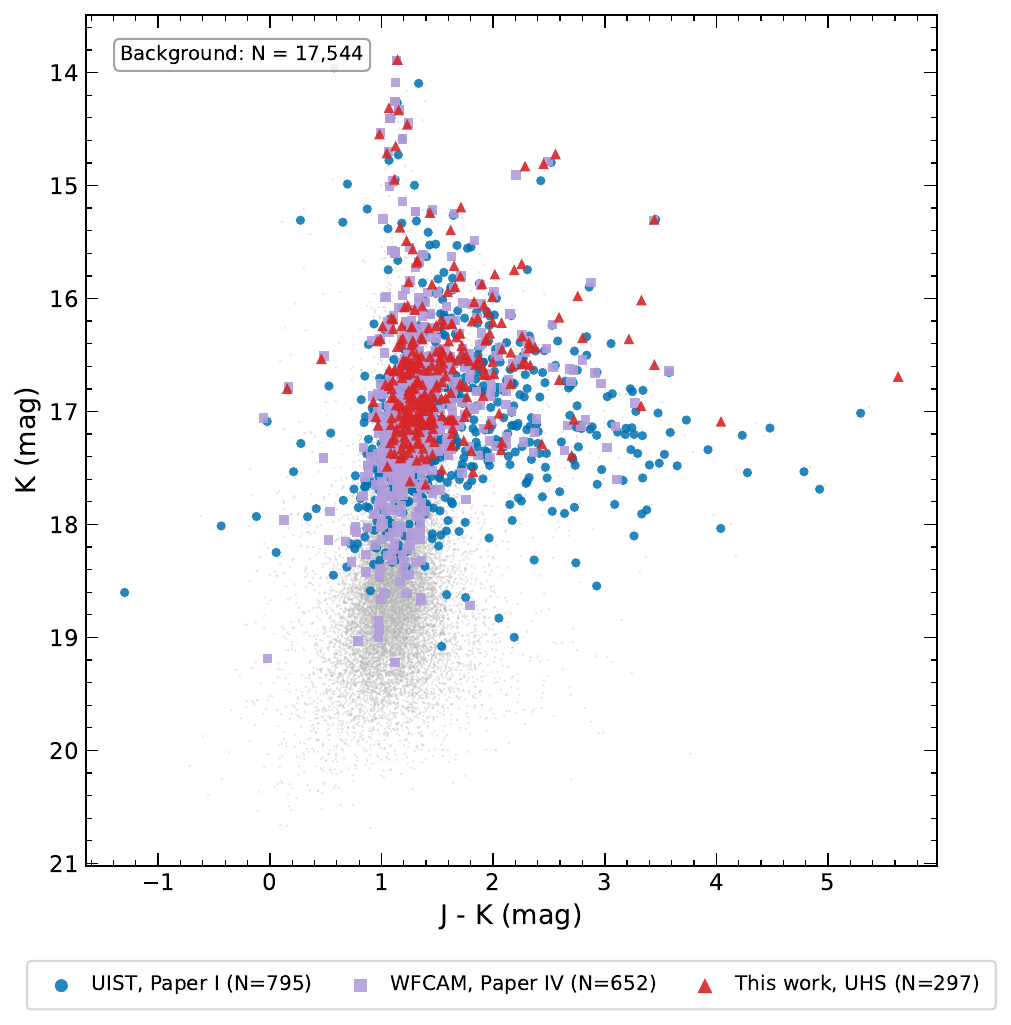}
\caption{Near-infrared colour--magnitude diagram of the central
kiloparsec of M\,33. Grey points show the full background population
of 17\,544 stars from \cite{Javadi2011a}.
Coloured symbols show variable stars from the three surveys: 795 of
the 812 UIST variables from Paper~I (circles; Fig.~\ref{fig:fields_iau}),
652 WFCAM disc variables from Paper~IV that fall within the central
3-arcmin field (squares; their $J,K$ photometry is taken from the
positionally-matched \cite{Javadi2011a}
catalogue, since the Paper~IV variable list itself provides positions
and a variability index only), and 297 of the 308 UHS variables from
this work (triangles). The 17, 204 and 11 stars missing from the
UIST, WFCAM and UHS samples respectively lack the $J$ photometry
needed to appear on this diagram.}
\label{fig:cmd_iau}
\end{center}
\end{figure}

\section{Future work}

Papers~VII, \cite{AlizadehVII}, and VIII defer two analyses that Paper~VII's extended
baseline enables for the first time.

\textbf{Star formation history.} The 385 high-amplitude Mira
candidates ($A_K \geq 1.0$~mag) among our variables have pulsation
periods that trace stellar birth mass through the
period--luminosity--age relation established for AGB stars by
\cite{Wood1999}. Robust individual periods,
unreliable from the fragmented $\sim$4-yr UIST-only baseline of
Papers~I--III, are now recoverable from our homogenised 15.11-yr
light curves. The forthcoming paper will measure these periods,
place each star on the period--luminosity relation, assign birth
masses via stellar evolutionary tracks, and reconstruct the star
formation history of the central kiloparsec, following the method
established for this galaxy by \cite{Javadi2011b} and applied across
its disc by \cite{Javadi2017}. The same LPV-based method has subsequently
been applied to the Large and Small Magellanic Clouds
\cite{Rezaeikh2014}, NGC~147 and NGC~185
\cite{HamedaniGolshan2017}, IC~1613 \cite{Hashemi2019}, M31
\cite{Torki2019,Torki2023,Torki2023b}, Andromeda~VII
\cite{Navabi2021}, Andromeda~I \cite{Saremi2021},  Andromeda~IX
\cite{Abdollahi2023}, IC~10 \cite{Gholami2023}, the halo of
NGC~5128 (Centaurus~A) \cite{Aghdam2024}, and NGC~6822
\cite{Khatamsaz2024}.

\textbf{Dust and mass return.} \cite{Javadi2013} first derived
mass-loss rates for AGB stars and
red supergiants in the M\,33 centre via SED modelling and
near-/mid-IR colour relations (see \cite{vanLoon2025} for a recent
review of red supergiant mass
loss); \cite{javadi2026ukinfraredtelescopem33} extended this across
the
full disc, finding that mass loss scales primarily with luminosity
and, more weakly, with pulsation period and amplitude. The total
mass-return rate to the interstellar medium, $\sim$0.1~M$_\odot$~yr$^{-1}$,
is about four times lower than the star formation rate, implying
that M\,33 requires external gas accretion to sustain star formation
beyond about a Gyr. Paper~VII already identifies 120 dust-enshrouded
AGB candidates ($K-[3.6]>0.5$~mag) among 156 Spitzer-matched stars in
the central field. The forthcoming paper will repeat this analysis
with individually-determined periods and amplitudes from the
15.11-yr baseline, yielding refined per-star mass-loss rates for the
central kiloparsec.

\end{document}